\documentclass[aps,preprint,prb,nofootinbib]{revtex4}
\usepackage[utf8x]{inputenc}
\usepackage{graphicx}
\usepackage{subcaption}
\usepackage{gensymb}
\usepackage{amsmath}
\usepackage{graphics}
\usepackage{float}
\usepackage{epsfig}
\usepackage{tikz}
\usepackage{caption}

\usetikzlibrary{decorations.pathmorphing}

\begin{document}

\title{On the Invariance of the Spacetime Interval}

\author{M.~Moriconi}
\email{mmoriconi@id.uff.br}
\affiliation{Instituto de Física, Universidade Federal Fluminense,
Campus da Praia Vermelha, Niterói, 24210-340, RJ, Brazil.}

\begin{abstract}
We present a geometric proof of the invariance of the relativistic spacetime interval based solely on the constancy of the speed of light, and the homogeneity and isotropy of spacetime. The derivation is based on a simple construction involving light rectangles, whose areas remain invariant across inertial frames. Based on this construction, we also derive the Lorentz transformations.
\end{abstract}

\maketitle

\section{Introduction}

Einstein's special theory of relativity is based on the principle that the laws of physics have the same form in all inertial frames of reference. This is the first postulate of the special theory of relativity (see Taylor and Wheeler's Spacetime Physics \cite{wheeler}, for example), and it was known since Galileo for mechanical phenomena. What was radical in Einstein's proposal was the extension of this principle to {\em all} laws of physics. Exploring the consequences of this extension led to the special theory of relativity. What was at first a seemingly innocuous statement revolutionized our understanding of the causal structure of space and time, challenging long held beliefs such as the idea of absolute space and time or absolute simultaneity.  The second postulate Einstein put forward originally was the constancy of the speed of light, $c$. It is interesting to note that this is a consequence of the first postulate, as Einstein himself later remarked in a footnote in his paper on the relation between inertia and energy \cite{emc2}: by considering the laws of electromagnetism as fundamental laws of nature, one is led to the fact that the speed of light in vacuum is the same in all inertial frames of reference.

In 1908 Hermann Minkowski \cite{Galison} provided a geometric picture of special relativity, introducing the four-dimensional spacetime continuum. Initially Einstein was not very fond of this geometric description of the special theory of relativity \cite{Pais}, but later he understood its value and embraced it thoroughly, turning it into one of the main conceptual and mathematical tools to develop the general theory of relativity.

One of the main novelties in this geometric view of relativity is the introduction of the spacetime interval as the four-dimensional distance between events. The spacetime coordinates of a physical event are given by $(ct,x,y,z)$, the time the event occurred together with its position in space, according to an arbitrary coordinate system. We use $ct$ for the ``time" coordinate so that it has the same dimension as space coordinates. Since the speed of light, $c$, is the same for all inertial observers, this ``calibration" of the time coordinate can be universally accepted. 

Suppose the spacetime coordinates of two physical events, measured in a given inertial frame of reference, are given by $(ct_1, x_1, y_1, z_1)$ and $(c t_2, x_2, y_2, z_2)$, respectively. The relativistic spacetime interval is defined to be
\begin{equation}
    (\Delta S)^2 = - c^2 (\Delta t)^2 + (\Delta x)^2 + (\Delta y)^2 + (\Delta z)^2 \label{interval}
\end{equation}
where $\Delta t = t_2 - t_1$ and so on. If $(\Delta S)^2 < 0$ we call it a timelike interval; if $(\Delta S)^2 = 0$ a lightlike (or null) interval; if $(\Delta S)^2 > 0$ a spacelike interval.

It is a fundamental result in relativity that the spacetime interval is invariant under the change of inertial frames of reference. There are essentially two main approaches to prove its invariance: either one finds the transformations that relate spacetime coordinates between different inertial frames and then discovers that $(\Delta S)^2$ as defined in Equation (\ref{interval}) is invariant under these transformations, or else one can prove $(\Delta S)^2$ is an invariant based on general principles and then, as a by-product, find the space and time transformations that leave it invariant.

In this paper we take this second approach, which has also been used by a number of authors. In presentations of relativity these go from the elementary, for example, Taylor and Wheeler's book \cite{wheeler}, to more advanced, such as Schutz \cite{schutz}. A very useful resource letter by Wald \cite{wald} states: ``All features of spacetime structure in special relativity can be derived from the spacetime interval".

We begin by assuming that spacetime is homogeneous and isotropic; that is, all points in space and time are equivalent (homogeneity) as well as all directions of space and time (isotropy). These assumptions imply that the transformations between the spacetime coordinates of a given physical event as recorded in two different inertial frames must be linear in the spacetime coordinates. See Levy-Leblond \cite{levyleblond} for more details.

Starting with these assumptions, proving the invariance of lightlike intervals is immediate, and it will be done at the beginning of Section II. The invariance in the other two cases is harder, and requires more work. Taylor and  Wheeler, for example, established the invariance of timelike intervals by means of a thought experiment involving mirrors in a train, but the argument cannot be extended to spacelike intervals. In Section II we show the invariance of timelike and spacelike intervals by means of a geometric construction using light rays. In section III we use this invariance to derive the Lorentz transformations and in Section IV we present our conclusions.

\section{Invariance of the spacetime interval}

Consider two inertial frames of reference {$\cal S$} and {$\cal S'$} in the standard configuration; that is, the respective spatial axis are parallel, their origins coincide at $t = t' = 0$, and the origin of {$\cal S'$} moves with velocity $v$ in the $+x$ direction with respect to ${\cal S}$. Throughout this work, symbols marked with a prime will represent measurements in ${\cal S'}$, and symbols without a prime will represent measurements of corresponding quantities in ${\cal S}$. 

Due to the isotropy of space, measurements of lengths orthogonal to the velocity give  the same results in {$\cal S$} and {$\cal S'$}. This can be understood in the following way: imagine a cylindrical tube with circular cross-section of radius $R$ in its own rest frame, whose velocity is parallel to its axis of symmetry, moving towards a circular hole in a wall with the same radius $R$ in the wall's rest frame. If perpendicular lengths were shortened, then observers in the wall's frame of reference would expect the ``moving" tube to have a smaller cross section than the hole so that it would pass through with room to spare. On the other hand, in the frame of reference of the tube, the hole would have a smaller radius, implying in a collision between the tube and the wall. A similar contradiction is found if perpendicular lengths increase. The only paradox-free case is to have perpendicular lengths invariant.

Now consider a few facts that follow from the invariance of the speed of light. First, the spacetime diagrams of light beams emitted from the origin in $\cal S$ at $t=0$ along the $\pm x$ axis are similar to their description in ${\cal S'}$, in that  they are both straight lines at $\pm 45$ degrees in each frame of reference, and not the usual ``slanted" version seen in textbooks (see, for example, Schutz\cite{schutz}), where one draws the spacetime axis of one frame $({\cal S}')$ in the other (${\cal S}$).  Figure \ref{Events_in_S} is the representation of two light rays emitted from $x =0$ at $t=0$, event ${\cal O}$, in which two light rays are emitted from the origin in opposite directions at $t = 0$ and their detection, events ${\cal L}$ to the left and ${\cal R}$ to the right. The propagation of the same light rays in ${\cal S'}$ would still be represented by straight lines at ±45°, although the length of each line might differ from its length in ${\cal S}$. 

\begin{figure}[H]
\centering
\begin{minipage}{0.3\textwidth}
\centering
\begin{tikzpicture}
    \draw[->] (-2,0) -- (2,0) node[below] {$x$};
    \draw[->] (0,-1) -- (0,2.5) node[left] {$ct$};
    \draw (0,0) node[below left] {$\cal O$} -- (1.5,1.5) node[above right] {$\cal R$};
    \fill (0,0) circle (2pt);
    \fill (1.5,1.5) circle (2pt);
    \draw (0,0) -- (-1.0,1.0) node[above left] {$\cal L$};
    \fill (-1.0,1.0) circle (2pt);
\end{tikzpicture}
\caption*{(a)}
\label{Events_in_S}
\end{minipage}
\begin{minipage}{0.3\textwidth}
\centering
\begin{tikzpicture}
    \draw[->] (-2,0) -- (2,0) node[below] {$x'$};
    \draw[->] (0,-1) -- (0,2.5) node[left] {$ct'$};
    \draw (0,0) node[below left] {$\cal O$} -- (1.0,1.0) node[above right] {$\cal R$};
    \fill (1.0,1.0) circle (2pt);
    \fill (0,0) circle (2pt);
    \draw (0,0) -- (-1.5,1.5) node[above left] {$\cal L$};
    \fill (-1.5,1.5) circle (2pt);
\end{tikzpicture}
\caption*{(b)}
\label{Events_in_S_prime}
\end{minipage}
\caption{Left and right moving rays observed in (a) the ${\cal S}$ frame of reference, and (b) the ${\cal S'}$ frame of reference. This figure shows that the lengths may be different in the two reference frames but the paths will still make $45^{\degree}$ angles to the axes.}
    \label{Events_in_S}
\end{figure}

The invariance of the speed of light also leads to the conclusion that lightlike intervals are invariant under transformation between reference frames; two events that can be connected by a light flash emitted from one and absorbed by the other have $\Delta x = c \Delta t$ in any inertial frame of reference, implying $\Delta S^2 = 0$.

Next, we establish the invariance of timelike intervals. Due to homogeneity of space and time, we can chose the origin so that one of the events has coordinates $(0,0)$ and the other has coordinates $(ct,x)$ with $ct > x$, so that  the event is located inside the light-cone with positive $t$.

Suppose we turn on a light bulb located at the origin of ${\cal S}$ at time $t = 0$, and that there are mirrors to the right and to the left of the origin, located at $x = r$ and $x = -l$, respectively, with $r$ and $l$ positive. As shown in Figure \ref{light_rectangles} $(a)$, the event  ``turning on a light bulb" will be labeled event $\cal O$ (spacetime coordinates $(0,0)$); the events ``reflection by the left/right mirrors" will be labeled events $\cal L$/${\cal R}$ (spacetime coordinates $(r,r)$/$(l,-l)$); and the event ``meeting of the reflected rays" will be labeled event $\cal P$ (spacetime coordinates $(ct, x)$). Since, as vectors, ${\cal OR} + {\cal OL} = {\cal OP}$, we have $(ct,x) = (r+l, r-l)$. In order to locate an arbitrary event ${\cal P}$ at any position inside the light cone, we can locate the mirrors at $r$ and $-l$ at:
\begin{eqnarray}
    r &=& (ct + x)/2 , \label{eq_for_r}\\
    -l &=& (- ct + x)/2. \label{eq_for_l}
\end{eqnarray}
We call this construction a ``light rectangle".

\begin{figure}[H]
\centering
\begin{minipage}{0.3\textwidth}
\centering
\begin{tikzpicture}
    \draw[->] (-2,0) -- (2,0) node[below] {$x$};
    \draw[->] (0,-1) -- (0,2.5) node[left] {$ct$};
    \draw (0,0) node[below left] {$\cal O$} -- (1.5,1.5) node[above right] {$\cal R$};
    \fill (0,0) circle (2pt);
    \fill (1.5,1.5) circle (2pt);
    \draw (0,0) -- (-1.0,1.0) node[above left] {$\cal L$};
    \fill (-1.0,1.0) circle (2pt);
    \draw (-1.0,1.0) -- (0.5,2.5) node[above right] {$\cal P$};
    \fill (0.5,2.5) circle (2pt);
    \draw (1.5,1.5) -- (0.5,2.5);
    \draw[dashed] (0,0) -- (0.5,2.5); 
\end{tikzpicture}
\caption*{(a)} 
\end{minipage}
\begin{minipage}{0.3\textwidth}
\centering
\begin{tikzpicture}
    \draw[->] (-2,0) -- (2,0) node[below] {$x'$};
    \draw[->] (0,-1) -- (0,2.5) node[right] {$ct'$};
    \draw (0,0) node[below left] {$\cal O$} -- (1.0,1.0) node[above right] {$\cal R$};
    \fill (1.0,1.0) circle (2pt);
    \fill (0,0) circle (2pt);
    \draw (0,0) -- (-1.5,1.5) node[above left] {$\cal L$};
    \fill (-1.5,1.5) circle (2pt);
    \draw (1.0,1.0) -- (-0.5,2.5) node[above left] {$\cal P$};
    \fill (-0.5,2.5) circle (2pt);
    \draw (-1.5,1.5) -- (-0.5,2.5);
    \draw[dashed] (0,0) -- (-0.5,2.5); 
\end{tikzpicture}
\caption*{(b)}
\end{minipage}

\caption{Light rectangle associated with the event $\cal P$ in (a) $\cal S$; and (b) $\cal S'$. The sides are all at $45^{\degree}$ to the axes because they represent the paths of light beams.}
\label{light_rectangles}
\end{figure}

Let us analyze, now, the relation between the spacetime coordinates of events $\cal L$ and $\cal R$ as described in ${\cal S}$ and ${\cal S}'$. The spacetime coordinates of the ${\cal L}$ and ${\cal R}$ events, as measured in $\cal S'$, are given by $(l',-l')$ and $(r',r')$, respectively. 

Due to homogeneity of space there must be a linear relation between the spacetime coordinates in $\cal S$ and $\cal S'$. To understand why this is so, consider the event $\cal R$, which occurred at position $r$ and time $t = r/c$. The corresponding coordinates in $\cal S'$ are position $r'$ and time $t' = r'/c$. Imagine we double the distance $r$. Since we could consider this a sequence of two events, first light going from the origin to $r$, then going from $r$ to $2r$, and since due to homogeneity of space we can take any point as the origin of our system of coordinates, the corresponding space coordinate in $\cal S'$ must be $r' + r' = 2 r'$, that is, if we double $r$, we double $r'$ \footnote{Since we can always think of a given position as a composition of many infinitesimal space steps, it is not difficult to see that if $r \to \alpha r$ than $r' \to \alpha r'$.}. Therefore we can write $r' = K_+ r$. Since event $\cal R$ is the detection of a light pulse emitted from the origin, the time coordinate in $\cal S'$ is $r' = K_+ r$ . Clearly this is also true for event $\cal L$, but since it is in the negative $x$ axis, we must allow for a different proportionality constant, $l' = K_- l$. Notice that $K_\pm$ may depend on velocity.  See the books by Taylor and Wheeler \cite{wheeler} or Schutz \cite{schutz} or the paper by Levy-Leblond \cite{levyleblond} for more formal discussions.

Assuming that $\cal S'$ moves with a velocity $v < c$ \footnote{It is impossible to start from rest and reach speeds greater than $c$, for this would contradict the constancy of the speed of light. An observer capable of exceeding $c$ could overtake a  pulse they sent forward and, at some point, see it propagating backward, in violation of relativity postulates.}, the event $\cal R$ ($\cal L$) has $r' > 0$ ($l' < 0$), events in front of the origin remain in front, events behind the origin remain behind, implying $K_\pm > 0$.

Further, we must have $r = K_- r'$. The coefficient is $K_-$, not $K_+$, because, for an observer in ${\cal S'}$, the light beam is propagating in the opposite direction as the velocity of the frame of reference ${\cal S}$. This, in turn, implies $r' = K_+ r = K_+ K_- r'$, that is, $K_+K_- = 1$. We can give this a geometric interpretation: {\em the areas \footnote{The area is actually $\sqrt{2} r \sqrt{2}l = 2rl$, but the factor $2$ is irrelevant for our argument} of the light rectangles are independent of the reference frame}, since $r'l' = K_+rK_-l = rl$. Using Equations (\ref{eq_for_r})  and (\ref{eq_for_l}), we obtain $rl = ( c^2  t^2 - x^2)/4 = - \Delta {\cal S}^2/4$, establishing the invariance of timelike intervals.

Notice that having the light pulse start at $(0,0)$ was just a convenience, and if it starts at an arbitrary position and time, we obtain as result the invariance of $\Delta S^2$ as given in Equation (\ref{interval}).

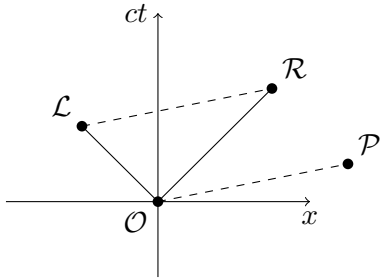
\begin{figure}[H]
\begin{center}
\begin{tikzpicture}
    \draw[->] (-2,0) -- (2,0) node[below] {$x$};
    \draw[->] (0,-1) -- (0,2.5) node[left] {$ct$};
    \draw (0,0) node[below left] {$\cal O$} -- (1.5,1.5) node[above right] {$\cal R$};
     \fill (0,0) circle (2pt);
    \fill (1.5,1.5) circle (2pt);
    \draw (0,0)  -- (- 1.0,1.0) node[above left] {$\cal L$};
    \fill (- 1.0,1.0) circle (2pt);
    \draw[dashed] (- 1.0,1.0) -- (1.5,1.5); 
    \draw[dashed] (0.0,0.0) -- (2.5,0.5) node[above right] {$\cal P$}; 
    \fill (2.5,0.5) circle (2pt);
\end{tikzpicture}
    \caption{How to translate an arbitrary spacelike vector ${\cal O P}$ to the vector ${\cal LR}$.}
  \label{spacelike_vector}
\end{center}
\end{figure}

We showed that timelike intervals are invariant by considering one diagonal of the light rectangle. We will show now that spacelike intervals are invariant by considering the other diagonal. Recall that an interval that starts at the origin is timelike if it falls inside the light cone and spacelike if it falls outside the light cone. We start by noting that any two events separated by a spacelike interval can be relabeled as a vector ${\cal LR}$ for $0 < l < \infty$ and $0 < r < \infty$ \footnote{We can always adjust the system of coordinates in such a way that the spatial coordinate of event $\cal P$ is positive.} by a translation of the origin of $x$ and $t$. Here we are using the homogeneity of spacetime: any point in spacetime can serve as the origin of our system of coordinates. In Figure \ref{spacelike_vector} we show an example of how to do this: the spacelike vector ${\cal O \cal P}$ can be translated to the vector ${\cal L \cal R}$. In this example, homogeneity of spacetime implies that the relativistic interval between events $\cal O$ ad $\cal P$ is the same as between events $\cal L$ and $\cal R$.

The spacetime interval between events $\cal L$ and $\cal R$ is $\Delta S^2 = - (r-l)^2 + (r + l)^2 =  4 rl$, which is proportional to the area of the light rectangle, implying the invariance of spacelike intervals. Together with the previous results we have established the invariance of any type of interval, be it lightlike, timelike or spacelike.

\section{Lorentz Transformations}

The Lorentz transformation is the precise mathematical relation between the spacetime coordinates of a given physical event as described in two different inertial frames of reference. Due to its geometric approach, we should mention a derivation of the Lorentz transformation by Macdonald \cite{macdonald} , which uses the constancy of the speed of light as one of the main ingredients of the derivation.

Initially, since $K_+ K_- = 1$, and $K_\pm$ is positive, we can write $K_\pm = e^{\pm \phi}$, with $-\infty < \phi < \infty$. First let us discuss the Lorentz transformation for timelike events.

We have $(ct, x) = (r+l, r-l)$ and $(ct', x') = (r' + l', r' - l')$, and, after some simple algebra and using Equations (\ref{eq_for_r}) and (\ref{eq_for_l}), we obtain
\begin{eqnarray}
    &&ct' = r' + l' = K_+ r + K_- l = ct \cosh \phi + x \sinh \phi  \nonumber \\
    &&x' = r' - l' = K_+ r - K_- l = ct \sinh \phi +  x \cosh \phi . \label{protoLorentz}
\end{eqnarray}
In order to fix the parameter $\phi$ notice that, according to an inertial observer in ${\cal S}'$, the spatial origin of $\cal S$ moves with velocity $-v$, implying $x' = - vt' = - \beta ct'$, where $\beta = v/c$. Since in this case the transformations (\ref{protoLorentz}) give $x' = ct \sinh \phi$ and $ct' = ct \cosh \phi$, we must have $\tanh \phi = - \beta$ and $\cosh \phi = 1/\sqrt{1 - \beta^2} \equiv \gamma$, giving, finally,
\begin{eqnarray}
    &&ct' = \gamma (ct - \beta x)  \nonumber \\
    &&x' = \gamma (x - \beta ct) . \label{LT}
\end{eqnarray}

We can treat spacelike events in the same fashion.
As vectors, from Figure \ref{spacelike_vector} we have ${\cal OL} + {\cal OP} = {\cal OR}$, implying that the coordinates of $\cal P$ are $(r-l, r+l)$. By the same reasoning, in $\cal S'$ its coordinates are $(r' - l',r' + l')$. Proceeding as in the timelike case, we obtain the same transformation for the coordinates of spacelike events, Equations (\ref{LT}). A similar analysis works as well for lightlike events.

This is the Lorentz transformation for the one-dimensional motion. Together with the remark that lengths perpendicular to the direction of velocity are the same in both frames, we have a fairly complete and useful description on how to obtain the spacetime coordinates of a physical event in different inertial frames.

\section{Conclusions}

While the invariance of the spacetime interval is often introduced via algebraic methods, as a consequence of the Lorentz transformations, here we have taken the opposite view, deriving the invariance of the spacetime interval first and then obtaining the Lorentz transformations. Establishing this invariance, especially for timelike and spacelike intervals, can be subtle. For instance, the discussion in Landau and Lifshitz \cite{landau} offers an intuitively appealing argument starting from lightlike intervals but, as shown by Elton \cite{elton}, it lacks full mathematical rigor.

It is also worth contrasting our approach with the axiomatic treatment initiated by Lévy-Leblond \cite{levyleblond}, and further developed by Lee and Kalotas \cite{Lee}, Ignatowsky \cite{Ignatowsky}, and Torretti \cite{Torretti}. Their framework explores the most general class of transformations between inertial frames that are compatible with basic spacetime symmetries, such as homogeneity, isotropy, causality, and group structure, without assuming the invariance of the speed of light. A more mathematical approach worth mentioning is the one by Zeeman \cite{zeeman}, where it is shown that causality alone implies that the Lorentz group is the symmetry group of the Minkowski space.

We presented a derivation of the invariance of the spacetime interval and the Lorentz transformation based on the constancy of the speed of light, along with the homogeneity and isotropy of spacetime. Our argument is essentially geometric, relying on the invariance of the area of light rectangles.

\begin{acknowledgments}

The author would like to thank the anonymous referees, whose comments helped improve this paper considerably. The author has no conflicts to disclose.

\end{acknowledgments}

\end{document}